# Altermagnetism and Room-Temperature Metal-to-Insulator Transition in CsCr$_2$S$_2$O

Yi Liu[*†1,2], Chen-Chao Xu[*3], Jin-Ke Bao[*3], Bai-Jiang Lv[*4,5], Hao Li[4,5], Jing Li[2], Yi-Qiang Lin[2], Hua-Xun Li[2], Yi-Ming Lu[2], Xin-Yu Zhao[2], Wu-Zhang Yang[2], Zhen-Yi Zhang[6], Xian-Yan Chen[7], Wen-he Jiao[1], Ji-Yong Liu[8], Bai-Ren Zhu[1], and Guang-Han Cao[†2,9,10]

[1]School of Physics, Key Laboratory of Quantum Precision Measurement of Zhejiang Province, Zhejiang University of Technology, Hangzhou 310023, China

[2]School of Physics, Zhejiang University, Hangzhou 310058, China

[3]School of Physics and Hangzhou Key Laboratory of Quantum Matters, Hangzhou Normal University, Hangzhou 311121, China

[4]Institute of Nuclear Physics and Chemistry, Chinese Academy of Engineering Physics, Mianyang 621999, China

[5]National Key Laboratory of Neutron Science and Technology, Mianyang 621999, China

[6]Bruker Scientific Instruments Co., Ltd, Shanghai 201100, China

[7]Material Scientific Cores, Zhejiang Laboratory, Hangzhou 311100, China

[8]Department of Chemistry, Zhejiang University, Hangzhou 310058, China

[9]Interdisciplinary Center for Quantum Information, and State Key Laboratory of Silicon and Advanced Semiconductor Materials, Zhejiang University, Hangzhou 310058, China

[10]Collaborative Innovation Centre of Advanced Microstructures, Nanjing University, Nanjing 210093, China,

*These authors contribute equally to this work.

†E-mail: liuyiphy@zjut.edu.cn; ghcao@zju.edu.cn

**Abstract**

Metal-to-insulator transitions (MITs), particularly near room temperature, have been extensively studied in nonmagnetic and conventional ferromagnetic and antiferromagnetic systems, yet the co-emergence of MIT and altermagnetism (AM) remains unexplored. Here, a layered chromium-based compound CsCr$_2$S$_2$O that realizes this coexistence was synthesized. It crystallizes in CeCr$_2$Si$_2$C-type structure with Cr moments orders in a C-type antiferromagnetic configuration below $T_\mathrm{N}$ = 326 K, constituting a room-temperature d-wave altermagnet. In the altermagnetic state, a subsequent Verwey-type MIT appears at $T_\mathrm{MI}$ = 305 K, driven by a tetragonal-to-orthorhombic structural distortion and stripe charge ordering of Cr$^{2+}$/Cr$^{3+}$ ions, while maintaining its altermagnetic character. First-principles calculations show moment-dependent spin-split electronic structures with maximum splitting energies of ~0.6 eV and ~0.3 eV in the metallic and insulating states, respectively. Our work links the two prominent phenomena, MIT and AM, in a single material, establishing a new platform for potential spintronic applications.

# 1 Introduction

Metal-insulator transitions (MITs) constitute a fundamental phenomenon in condensed matter physics with profound implications for next-generation electronic devices, including ultrafast switches and non-volatile memories [1]. They could arise from diverse mechanisms. In the absence of long-range magnetic order, the routes include Mott localization driven by strong electron correlations [2, 3], as exemplified by $Ca_{1.9}Sr_{0.1}RuO_4$ [4], and Peierls lattice instabilities involving dimerization such as $K_{0.3}MoO_3$ [5]. MITs can also emerge through intimate coupling to spin configurations, such as the Slater mechanism in antiferromagnets [6, 7], and spin-driven Lifshitz transitions [8].

Materials exhibiting both MIT and spin-split electronic bands are particularly promising for spintronics [9, 10, 11, 12]. Conventionally, such spin splitting is achieved through exchange interactions in ferromagnets or strong spin-orbit coupling (SOC) in heavy-element compounds. In manganese-based perovskite oxides, the MITs are coupled with ferromagnetic transitions, exhibiting colossal magnetoresistance arising from spin-polarized transport [13]. Similarly, $Fe_3O_4$ with Verwey-type MIT driven by $Fe^{2+}/Fe^{3+}$ charge order in a ferromagnetic background, enables spin-transport modulation [12]. Alternatively, in $4d$ and $5d$ transition-metal oxides with MITs, such as $Sr_2IrO_4$ [14], strong SOC can also induce spin splitting. However, ferromagnetic ordering introduces detrimental stray fields that compromise device scalability, whereas SOC requires high-Z elements and typically yield only modest spin-split energy.

In this context, the recent proposed altermagnetism (AM) offers a compelling alternative, which emerges as a distinct form of magnetic order combining the zero net magnetization of antiferromagnetism (FM) with the spin-split capability of ferromagnets [15, 16, 17, 18]. The spin splitting in AM is symmetry-protected and moment-dependent with typically large maximum splitting energy. To date, AM has been theoretically predicted in various materials [19, 20] and is increasingly supported by experimental evidence [21, 22, 23, 24], with candidates spanning both metals [21, 22, 23, 24, 25] and insulators [26, 22]. However, MIT occurs in the background of AM remains unexplored currently.

The $T_2OCh_2$ ($T$ = V, Cr, Fe; $Ch$ = S, Se, Te) motif within the $CeCr_2Si_2$-type (1221) structure has emerged as a promising platform for realizing altermagnetism (AM) [16, 27, 28, 29]. In this layered structure, nearest-neighboring magnetic $T$ atoms are coordinated by O and $Ch$ atoms in an asymmetric fashion, breaking both translation ($\tau$) and inversion ($I$) symmetries while preserving specific rotation or mirror symmetries. Previous studies initially hypothesized C-type antiferromagnetic (AFM) order for $KV_2Se_2O$ [30] and $Rb_{1-\delta}V_2Te_2O$ [31] to explain observed spin-split signatures. However, recent neutron diffraction investigations have unequivocally identified G-type AFM configurations in $KV_2Se_2O$ [32] and $Cs_{1-\delta}V_2Te_2O$ [33]. Unlike the C-type phase, the G-type structure features antiferromagnetic interlayer coupling that preserves the symmetry forbidding bulk altermagnetic spin splitting, thus complicating the interpretation of previous observations in vanadium-based 1221 systems.

Here, a new chromium-based 1221 compound, $CsCr_2S_2O$, was synthesized and comprehensively investigated. It exhibits a cascade of phase transitions with AFM order at $T_N$ = 326 K and MIT at $T_{MI}$ = 305 K. Altermagnetism is identified in the magnetic ordering state with C-type AFM configuration. This altermagnetic feature is preserved across the MIT, which is characterized as Verwey type occurring concomitantly with displacive structural modulation and stripe charge order. A local Coulomb repulsion $U$ = 2.5 eV is expected in density-functional-theory (DFT) calculations to open a gap of 200 meV, which is close to the value estimated by experiment. In the metallic altermagnetic state, momentum dependent spin-split $d_{xz}$ and $d_{yz}$ electrons dominate the Fermi surfaces in the metallic altermagnetic state with the other $3d$ orbitals pushed away from $E_F$ by correlation effect. This results in a extremely anisotropic polarization, enabling giant opposite-spin currents in perpendicular directions, which is crucial for spin transport. In the insulating altermagnetic phase, spin polarized $d_{xz}/d_{yz}$ electrons from $Cr^{2+}$ and $Cr^{3+}$ form nearly parallel bands below and above the gap, respectively. The spin-split energy reaches ~0.3 eV, showing great potential for achieving efficient spin-filtering [34]. This work establishes $CsCr_2S_2O$ as a rare platform for reversible switching between metallic and insulating altermagnet at room temperature.

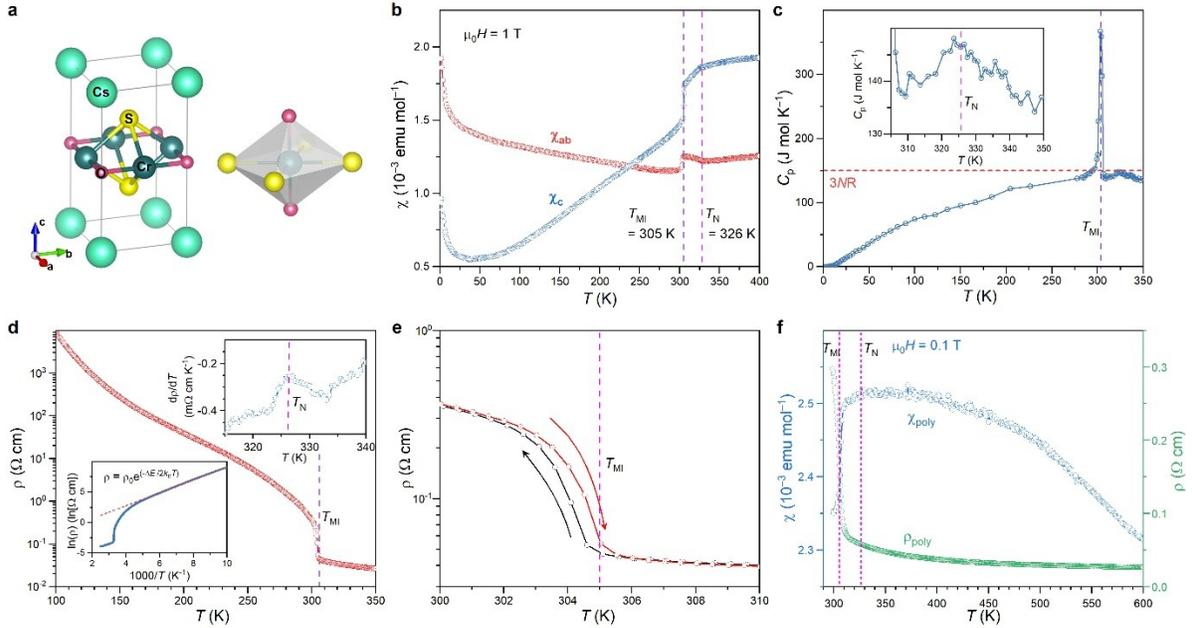

Figure 1: Crystal structure and physical properties of $CsCr_2S_2O$. a, High-temperature tetragonal crystal structure of $CsCr_2S_2O$(left) containing octahedral of $CrS_4O_2$(right). b, Anisotropic magnetic susceptibility with magnetic fields applied parallel ($\chi_{ab}$) and perpendicular ($\chi_c$) to the $ab$ plane. c, Temperature dependence of the specific heat. The inset is a close-up around the magnetic transition. d, Electrical resistivity as a function of temperature. The top right inset shows the temperature derivative of resistivity near the magnetic transition. The bottom left plots $\ln(\rho)$ versus $1/T$. The red dashed line represents the linear fit to extract the activation energy. e. Temperature dependence of resistivity in cooling and heating cycles. f, Resistivity and susceptibility measured on polycrystals from 300 K to 600 K.

## 2 Results

### 2.1 Characterization of $CsCr_2S_2O$

High-quality powdered and single-crystalline $CsCr_2S_2O$ samples were synthesized, as detailed in the Methods. The compositional and structural characterizations are summarized in Figure S1, including energy-dispersive X-ray spectroscopy (EDX), powder X-ray diffraction (PXRD) and single-crystal X-ray diffraction (SCXRD). EDX elemental analysis confirms molar ratios of Cs, Cr and S close to the nominal composition without detectable vacancies, and the deviation of O is attributed to adsorbed oxygen on the crystal surface. Refinement of SCXRD data collected at 310 K confirms that $CsCr_2S_2O$ crystallizes in space group $P4/mmm$ (No. 123), with detailed structural parameters listed in Table S1. It is constitute of alternated stacking of $Cr_2OS_2$ layer and Cs cation layer along the $c$ axis (Figure 1a), and Cr ion resides in an octahedral coordination environment, bonding to two O and four S atoms. Based on the refined bond lengths of Cr–S and Cr–O, the bond valence sum (BVS) [35] calculation for the Cr ion yields a value of +2.53, well consistent with the formal oxidation state of $Cr^{+2.5}$ in $CsCr_2S_2O$.

### 2.2 Phase transitions in $CsCr_2S_2O$

Figure 1b–f present the measurements of susceptibility ($\chi$), resistivity ($\rho$) and specific heat ($C_p$). The $\chi(T)$ curves display kinks at $T_N = 326$ K, with an upward kink in $\chi_{ab}$ and a downward kink in $\chi_c$. This anisotropy suggest that the magnetic moments of Cr are predominantly antiferromagnetically aligned along the $c$ axis,

which is further confirmed by the subsequent neutron powder diffractions (NPD). More susceptibility and magnetization data are presented in Figure S2. Notably, the ferromagnetism-like bifurcation of susceptibility measured in zero-field-cooling and field cooling processes under low fields, is attributed to the dangling chains at the naturally cleaved crystal edges, which also caused misinterpretation in $CoSe_2O_5$ [36]. The $T_N$ shows no detectable field dependence up to 5 T, demonstrating the robustness of the magnetic order against magnetic fields. Magnetization curves for fields parallel to the *ab* plane and *c* axis are both linear with no hysteresis, further supporting antiferromagnetism in $CsCr_2S_2O$. Around $T_N$, a subtle decrease of resistivity is identified by the peak in $d\rho/dT$ (top-right inset of Figure 1d). In the specific heat(Figure 1c), instead of distinct jump for a second-order magnetic transition, a broad peak kicks at $T_N$. High-temperature susceptibility measurement on polycrystalline samples up to 600 K shows a broad maximum around 400 K (Figure 1f). This feature suggests the existence short-range magnetic correlations persisting above $T_N$, which is also observed in similar systems [37, 38]. Indeed, the revealed C-type AFM order below demonstrates ferromagnetic align of the next-nearest neighboring Cr moments. This ferromagnetic tendency frustrates the antiferromagnetic superexchange interactions via linear Cr-O-Cr(180°) and bent Cr-S-Cr (103.14°) bridges as predicted by the Goodenough-Kanamori rules. Correspondingly, a significant portion of the magnetic entropy could be released gradually over a wide temperature range above $T_N$, which diminishes the specific-heat anomaly.

A successive transition happens at $T_{MI}$ = 305 K, marked by an abrupt surge in resistivity and a sharp drop in susceptibility. The resistivity increases more than one order of magnitude within a narrow temperature interval, establishing a electronic phase transition in $CsCr_2S_2O$. In the high-temperature (HT) phase, the system resides in a bad-metal regime characterized by a weakly negative temperature dependence of resistivity, a behavior typically associated with electron correlations and magnetic scattering. An Arrhenius fit to the data between 100 and 150 K reveals an activation gap of 196(2) meV (bottom-left inset of Figure 1c), confirming the insulating nature of the low-temperature (LT) phase. The first-order character of the phase transition from metal to insulator is corroborated by the thermal hysteresis in $\rho(T)$ and a prominent discontinuous spike in the specific heat.

## 2.3 Structure modulation in $CsCr_2S_2O$

To elucidate the structural evolution accompanying the electronic transitions, SCXRD below $T_{MI}$ was performed. As presented in the comparison of reconstructed reciprocal space maps (*hk*0) at 310 K and 180 K (Figure 2a), apart from the primary Bragg diffractions, systematic satellite reflections emerge at low temperature. These satellite spots can be interpreted by a modulated structure with a single commensurate propagation vector **q** = (0.5, 0.5, 0). No modulation along the *c* axis is identified. The intensities of the satellite reflections is more than one order of magnitude weaker than that of the main Bragg peaks. Leveraging the high-quality SCXRD data, we solved the modulated structure within a $\sqrt{2}a \times \sqrt{2}b \times c$ supercell along with a distortion from tetragonal to orthorhombic. The loss of fourfold rotational symmetry is corroborated by PXRD, which shows pronounced splitting of the (312) and (132) reflections below $T_{MI}$ (Figure 2e). To avoid ambiguity between phases, moment indices in the aLT phase are labeled with a subscript L. $(hkl)_L$ and $(hkl)$ are related by,

$$\begin{pmatrix} h \\ k \\ l \end{pmatrix}_L = \begin{pmatrix} 1 & 1 & 0 \\ -1 & 1 & 0 \\ 0 & 0 & 1 \end{pmatrix} \begin{pmatrix} h \\ k \\ l \end{pmatrix} \quad (1)$$

Table S2 lists the detailed crystallographic parameters for the refined structure with space group *Pmam* (No. 51). Corresponding reconstructed reciprocal space maps are shown in Figure S4a–c. While systematic absences are expected for reflections $(h0l)_L$ with odd *h*, the presence of a secondary twin domain rotated by 90° about the *c*-axis accounts for the observation of certain symmetry-forbidden reflections. Figure 2c shows the comparison of reconstructed planes$(h1l)_L$ and $(1kl)_L$, in which the intensities of forbidden $(101)_L$ and $(102)_L$ reflections are less than one-tenth those of $(011)_L$ and $(012)_L$ spots.

The LT crystal structure is depicted in Figure 2d-f. The Cr1–Cr1 and Cr1–Cr2 distances refine to 2.8142(6) Å and 2.8024(5) Å, respectively, corresponding to half the *a* and *b* lattice parameters.

Temperature-dependent lattice parameters extracted from PXRD reveals a contraction of Cr1–Cr1 bonds and elongation of Cr1–Cr2 bonds just below $T_{MI}$, suggesting the formation of Cr1 and Cr2 chains along the $a$ axis. The most distinctive feature of the insulating phase is the displacive modulation in which the Cs, S and O atoms undergo concerted lateral displacements along the $b$ axis. Specifically, the S atoms shift towards the Cr2 chains, resulting in Cr2–S bond distance 3.0% shorter than the average values. The position O atoms are slightly off-center to Cr2 atoms by 0.2%. Concurrently, the Cs atoms polymerize into zigzag chains positioned directly below or above Cr1 chains, with the minimum Cs–Cs distance contracted by 2% relative to the mean separation. BVS analysis yields oxidation states of +2.22 and +2.81 for Cr1 and Cr2, respectively, which are distinct from +2.53 at 310 K and closely approaching the formal charges of $Cr^{2+}$ and $Cr^{3+}$. This charge disproportionation indicates the formation of a stripe charge order coincident with the structural distortion, which drives the Verwey-type MIT, analogous to that proposed for $Fe_3O_4$ [39].

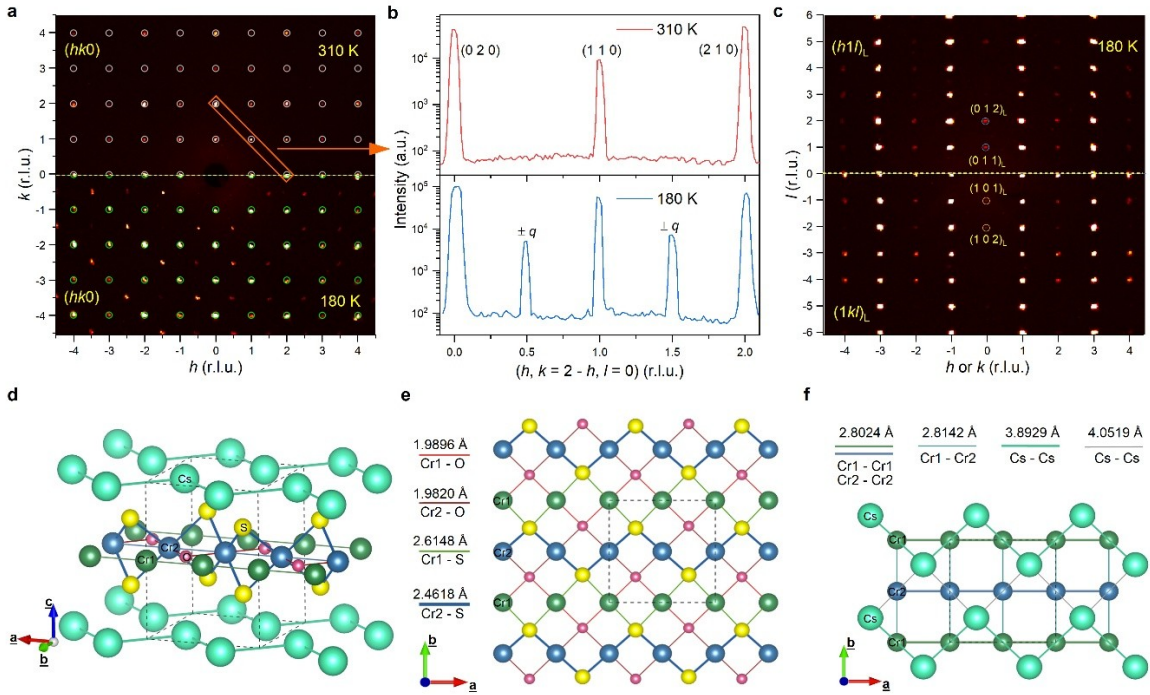

Figure 2: Structural modulation in $CsCr_2S_2O$. a, Comparison of reconstructed ($hk0$) planes of reflections at 310 K (top) and 180 K (bottom). b, Intensity of reflection cut from b at 310 K and 180 K. c, The ($h1l$) and ($1kl$) maps at 180 K. d. Low-temperature crystal structure of $CsCr_2S_2O$. e, f. Distorted structure of the $Cr_2S_2O$ and Cs layers.

## 2.4 Magnetic structure in $CsCr_2S_2O$

The antiferromagnetic order was investigated using neutron powder diffraction (NPD). Figure 3a compares normalized NPD patterns collected at 10 K, 315 K, and 500 K. The diffraction patterns at 500 K can be well fitted using the HT crystal structure with no magnetic order present. At 315 K, dramatic enhancements of reflections at 27.4° and 31.3° indicate the establishment of long-range magnetic ordering in $CsCr_2S_2O$. Therefore, Rietveld refinements of patterns below $T_N$ incorporate both nuclear and magnetic scattering contributions. Using the $k$ search program integrated in the FullProf suite, the magnetic propagation vector is identified as $k = 0$, indicating a commensurate magnetic structure with magnetic cell identical to the crystallographic lattice. Magnetic representation analysis based on $k = 0$ for space group

*P*4/*mmm* with SARAh package yields only two possible irreducible representations for the Cr sites: ferromagnetism and C-type AFM configuration. The latter is characterized by antiferromagnetic coupling between nearest-neighbor spins within *ab* plane and ferromagnetic spin alignment along *c* axis. As the magnetic scattering is only sensitive to the component of the moment perpendicular to *Q*, the prominent (100) magnetic reflection at 27.4° suggest the Cr moments aligned along the *c* axis(Figure 3b). consistent with the measurements of anisotropic susceptibility. The C-type antiferromagnetic model fits the diffraction pattern at 315 K quite well, yielding a moment of 1.82(4) $\mu_B$.

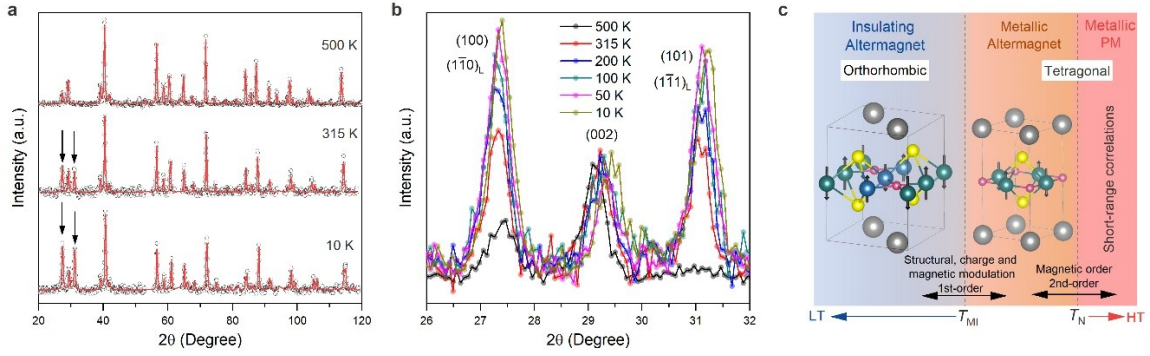

Figure 3: Magnetic structure of $CsCr_2S_2O$. a, NPD patterns at 500 K, 315 K and 10 K, and corresponding Rietveld refinements. b, The evolution of the primary peak from 10 K to 315 K. c. Magnetic structures of the HT and LT phase. Primary features in each phase are also summarized, where PM represents paramagnetism. The phase transition order and related phenomenon are also provided.

Upon cooling below $T_{MI}$, the magnetic peaks gradually increase without the emergence of new detectable peak, indicating that the magnetic configuration basically retains in the space group *Pmam*. Thus, the same arrangement of moments is used to fit the LT patterns. To ensure stable convergence in the refinements, the magnetic moments of Cr1 and Cr2 were initially constrained to be equal, despite the charge order of $Cr^{2+}$ and $Cr^{3+}$ implying a distinct spin moment of 4 $\mu_B$ and 3 $\mu_B$, respectively. Such a magnetic modulation in magnitude has been experimentally observed in $KV_2Se_2O$ by $^{51}V$ via nuclear magnetic resonance [30] and single-crystal neutron diffraction [32], occurring in the absence of structural modulation. As shown in Figure S5d, simulation reveals that satellite peaks associated with these magnetic modulation exhibit intensities merely 2% of the highest magnetic peak, which is too weak for unambiguous experimental detection here. The magnetic structures for HT and LT phases are summarized in Figure 3c, where the primary properties of each phases are also labelled. The refined magnetic moment of 2.65(5) $\mu_B$/Cr at 10 K is smaller than expected, likely owing to the unaccounted contributions from these weak satellites. The refinements for different temperatures are presented in Figure S5a. Notably, the temperature dependence of magnetic moment follows the power law $M_{Cr}(T) = C(1 − T/T_N)^\beta$ with $\beta = 0.11(3)$ with a more rapid arise just below $T_N$, rather than the Brillouin function predicted by mean-field theory(Figure S5b). The $\beta$ value is close to that of two-dimensional Ising model ($\beta = 0.125$), consistent with the quasi-two-dimensional nature of the crystal structure and the *c*-axis alignment of the moments.

## 2.5 Altermagnetism and electronic structure in $CsCr_2S_2O$

Based on the crystal and magnetic structures, we elucidate the symmetry origin of the altermagnetic spin splitting for HT and LT phases. As illustrated in Figure 4a, the HT phase hosts spin sublattices with opposite moments connected by the $[C_2\| C_{4z}]$ operations, or equivalently, $[C_2\| M_{110}]$ and $[C_2\| M_{1\bar{1}0}]$, instead of parity-time reversal or translation-time reversal symmetries. Consequently, alternating spin splitting is expected along the (100) and (010) directions in reciprocal space, as shown in Figure 4b. For the LT phase, the opposite spin sublattices at Cr1 or Cr2 sites are connected by nonsymmorphic *a*-glide involving mirror reflections, specifically $[C_2\| M_a| (1/2, 0, 0)]$ or $[C_2\| M_b| (1/2, 0, 0)]$ with mirror planes perpendicular to the *a*-axis or *b*-axis, repectively. As a result, altermagnetic spin splitting appears along the $(1\bar{1}0)_L$ and $(110)_L$ directions as shown in Figure 4d. Indeed, the (100) and (010) directions correspond to $(1\bar{1}0)_L$ and $(110)_L$, respectively, as Equation (1) defined, revealing that the spin-split pattern persists in real space across both metallic and insulating states. This non-uniform spin texture in the $ab$ plane is characterized as d-wave splitting[40].

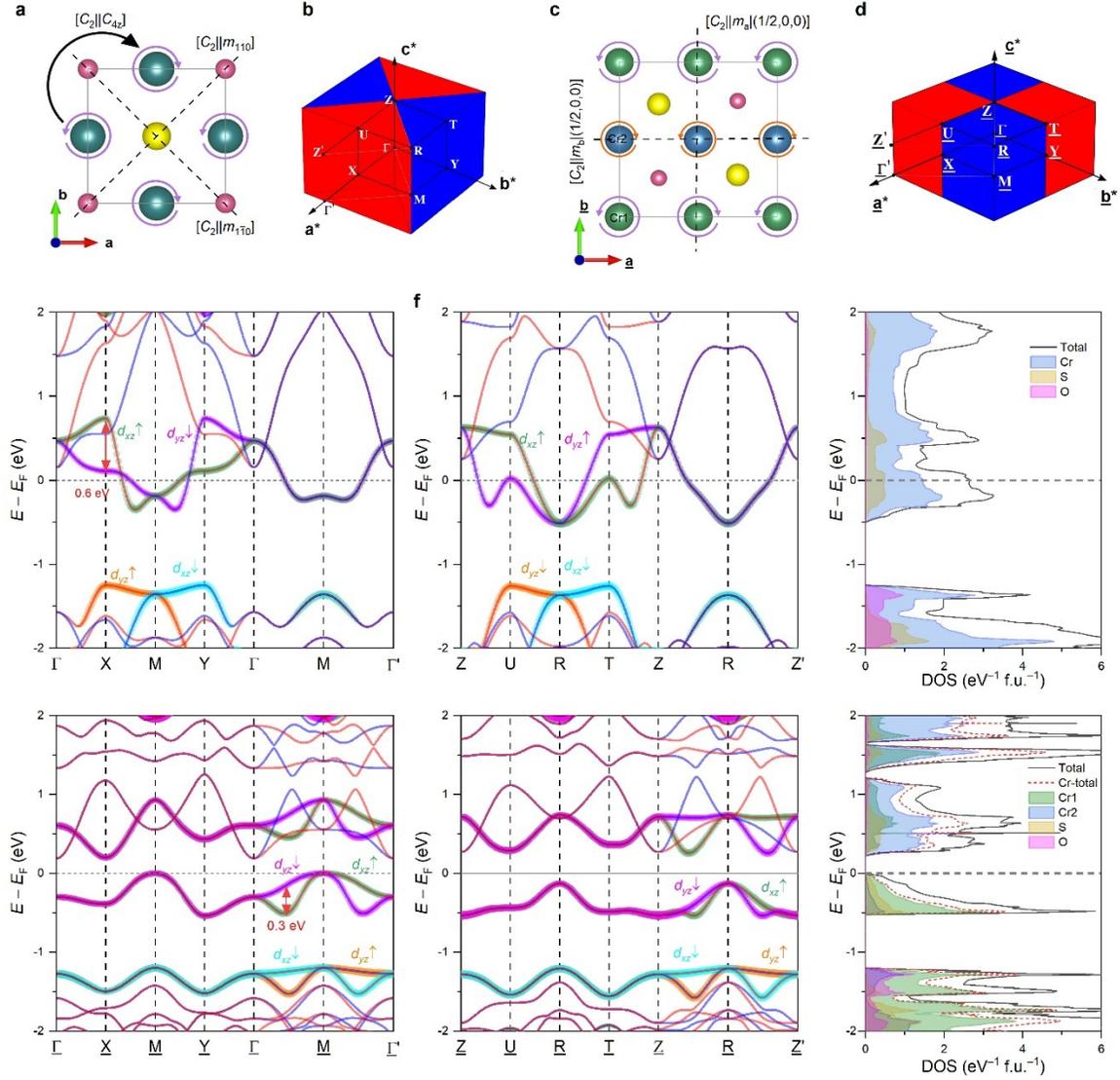

Figure 4: Symmetry analysis and electronic structure of CsCr$_2$S$_2$O. a,c, Top views of Cr$_2$S$_2$O layer for HT(left) and LT(right) phases. The arrows surrounding Cr atoms denote the up(anticlockwise) and down(clockwise) spins. The operation symmetries connecting the opposite-spin sublattices are labelled. b,d, Schematic illustration of the spin-splitting sign in the three-dimensional Brillouin zone for the HT(left) and LT(right) phases. e–j, Calculated electronic band structures and DOS for magnetic ordered HT and LT phases, using $U = 2.5$ eV and neglecting the spin-orbit coupling. For comparison with HT phase, the LT $d$-orbital are assigned the same orientation as the HT states in real space, equivalent to a 45° anticlockwise rotation around the $c$ axis.

First-principles calculations for HT and LT phases are further performed to elucidate the electronic landscape of the magnetically ordered states. Considering the opening energy gap below MIT, we firstly check the density-of-states (DOS) through systematic tuning of on-site Coulomb repulsion $U$ (Figure S6). In the AFM HT phase, the calculations show robust metallic state against $U$ with DOS at Fermi level($E_F$) nearly unchanged. While for the LT phase, a band gap starts to open at $U$ = 2 eV, and increases to 200 meV at $U$ = 2.5 eV, which matches well with the experimental value of 196 meV estimated from the resistivity. The $U$-dependence gap suggests that electronic correlation plays as an important role for the insulating behavior.

Based on $U$ = 2.5 eV, the band structures and DOS for HT and LT phases are calculated, as shown in Figure 4e-j. In the HT phase, the $E_F$ is raised by approximately 0.8 eV compared with analogous vanadium-based compounds, such as $CsV_2Te_2O$ [41]. Unlike in vanadium-based system, the Fermi level is dominated exclusively by Cr-$d_{xz}$ and Cr-$d_{yz}$ bands, with considerable hybridization from S atoms (Figure 4h), while the other orbital is pushed above $E_F$ by the local Coulomb repulsion, as evidenced through comparison with $U$ = 0 band structure in Figure S7. The AM related momentum-dependent spin polarization manifests as $d_{xz}$ and $d_{yz}$ bands splitting along the paths $\Gamma - X(Y) - M$ at $k_z$ = 0 and $Z - U$ $(T)-R$ at $k_z$ = 0.5, while remaining spin-degenerate along the $\Gamma - M$ and $Z - R$ lines. A polarization reverse appears between $Y$ and $M$. Additionally, the spin-down $d_{xz}$ (spin-up $d_{yz}$) and spin-up $d_{xz}$ (spin-down $d_{yz}$) are completely gaped by an energy of ~0.7 eV, indicating a full spin polarization of $d_{xz}$ and $d_{yz}$ electrons. The Fermi surface shows significant anisotropy (Figure S7), which will yield distinct group velocities in opposite-spin channels and thereby enable highly polarized currents vital for high-performance spintronics.

For the LT phase, the spin splitting is along the $\Gamma - M$ and $Z - R$, and degenerate in $\Gamma - X (Y)-M$ at and $Z - U (T)-R$. Therefore, the spin-split characters in HT and LT phase are preserved, consistent with the symmetry analysis. The spin-split energies reach about ~0.6 eV and ~0.3 eV for the metallic and insulating phase, respectively. The former are smaller than 1.6 eV reported in $KV_2Se_2O$[30], and the latter is comparable with insulating altermagnet MnTe[26]. For the tetragonal-to-orthorhombic transition, the bands at $X$ and $Y$ are asymmetric. Interestingly, two nearly parallel $d_{xz}(d_{yz})$ bands locate on opposite sides of the band gap. The DOS in Figure 4h shows that the lower bands are mainly contributed by Cr1, while the upper bands are dominated by Cr2. Additionally, calculations in the HT phase yields a local moment of ~3.2 $\mu_B$, aligning well with the 3.5 $\mu_B$ expected for $Cr^{2.5+}$, while for the LT phase Cr1 and Cr2 develop inequivalent moments of 3.5 $\mu_B$ and 2.9 $\mu_B$, respectively. These values are close to those expected for $Cr^{2+}$ (4 $\mu_B$) and $Cr^{3+}$ (3 $\mu_B$). The energy separation of Cr1/Cr2 charges and concomitant modulation of magnetic moments verify the formation of a charge-ordered state within the insulating phase.

## 3    Summary

Through systematic investigations of crystal, magnetic and electronic structures of the synthesized $CsCr_2S_2O$, we have demonstrated the co-emergence of AM and MIT near room temperature. This layered 1221 series exhibits C-type antiferromagnetic order below 326 K, forming a d-wave altermagnet with substantial momentum-dependent spin splitting up to ~ 0.6 eV. Upon cooling to 305 K, it undergoes a Verwey-type MIT driven by stripe charge ordering of mixed-valence $Cr^{2+}/Cr^{3+}$ ions and a structural transition from tetragonal to orthorhombic symmetry. Crucially, the spin-split electronic structure per-sists into the insulating phase with splitting energy of 0.3 eV, enabling a reversible switching between metallic and insulating altermagnet at ambient condition. Our discovery bridges correlated electron physics and spintronics, establishing a new class of multi-function quantum materials. The Verwey-type MIT transition in $CsCr_2S_2O$, which is driven by electronic charge ordering and lattice symmetry breaking, suggests a compelling avenue for tuning MIT through external stimuli such as strain[42] or electric fields[43, 44] in the future.

## 4    Methods

**Crystals Growth and Characterizations:** Single crystals of $CsCr_2S_2O$ were grown using CsCl as flux. Firstly, CrS was presynthesized by heating Cr and S powder at 750 °C in an evacuated silica ampoule. Secondly, the resulting CrS together with Cs, $Cr_2O_3$, S and CsCl in a molar ratio of Cs:CrS:$Cr_2O_3$:S:CsCl =

3:4:1:2:18, were loaded into an alumina crucible. This crucible was then sealed in a welded Ta tube, which was subsequently enclosed in a silica ampoule. The assembly was heated to 950 °C over 24 hours and held for 18 hours, followed by slow cooling to 600 °C at a rate of 2 °C/hour. Finally, the flux was removed by washing with deionized water, yielding black plate-like crystals with dimensions up to $1 \times 1 \times 0.2$ mm$^3$.

$CsCr_2S_2O$ polycrystals were synthesized by solid-state reactions. The source materials Cs, CrS, $Cr_2O_3$ and S were put into an alumina crucible subsequently in molar ratio and then sealed in an evacuated silica ampule. After prereaction at 500 °C for 16 hours, the sample was grounded into powder and then pressed into pellet in an argon-filled glove box. The pellet was sinter in evacuated silica ampoule at 700 °C for 24 hours. By repeating the above processes for 2–3 times, pure phase of $CsCr_2S_2O$ polycrystals were obtained.

Single-crystal XRD was carried on a Bruker D8 Venture diffractometer with Mo K$\alpha$ radiation. A piece of $CsCr_2S_2O$ crystal with dimensions of $0.145 \times 0.135 \times 0.026$ mm$^3$ was mounted on a MiTeGen loop. A flow of nitrogen gas was used to cool the crystal. A full data set was collected at each temperature. The data integration, scale and analysis were conducted on the commercial software packages APEX4, and structure solving and refining is done using Jana2020. The powder X-ray diffraction was conducted on a PANalytical X-ray diffractometer (Model EMPYREAN) with a monochromatic CuK$\alpha_1$ radiation, which was refined by Fullprof package. The chemical composition of the as-grown crystals was determined using EDX spectroscopy on a scanning electron microscope (Hitachi S-3700N) equipped with Oxford Instruments X-Max spectrometer.

**Physical Property Measurements:** The measurements of low-temperature electrical resistivity and specific heat were carried out on a physical property measurement system (PPMS-9, Quantum Design). The resistivity was measured by a standard four-terminal method. Specific heat was measured using thermal relaxation method with dozens of the crystals (total mass was 0.70 mg). The crystals were glued on the heat capacity puck with N grease. High-temperature resistivity up to 600 K was conducted on a commercial system (LSR-3, Linseis). The data of addenda were measured in advance. The magnetic measurements were performed on a magnetic property measurement system (MPMS-3, Quantum Design).

Neutron powder diffraction (NPD) experiments under different temperatures were performed using high-resolution neutron diffractometer at the Key Laboratory of Neutron Physics, Institute of Nuclear Physics and Chemistry, China Academy of Engineering Physics. Approximately 6 g of the $CsCr_2S_2O$ polycrystals were used for the measurement. The neutron wavelength is 1.8846 Å Scattering data were collected at various temperatures of 500, 315, 200, 100, 50 and 10 K. The collected diffraction patterns were refined via the Rietveld method using the Fullprof software [45].

**Electronic structure calculations:** Density functional theory (DFT) calculations were performed using VIENNA AB INITIO SIMULATION PACKAGE (VASP) [46, 47]. The Perdew, Burke, and Ernzerhof parametrization (PBE) of generalized gradient approximation to the exchange correlation functional was used [48]. Throughout the calculation, a plane-wave cutoff energy of 600 eV were adopted for the non-relativistic calculation. Using the lattice constants from experiment XRD data, the atom relaxation was performed with a Γ-centered k-point mesh of $12 \times 12 \times 6$ for the high-symmetry ($P4/mmm$) structure and $8 \times 8 \times 6$ for the low-symmetry ($Pmam$) structure, respectively. The relaxation reaches convergence until the force on each atom was less than 1 meV/Å and internal stress less than 0.1 kbar. The band structure and density of state calculation were based on optimized lattice structure. A tight-binding (TB) model Hamiltonian was fitted from the DFT results using the maximally localized Wannier functions (MLWF) method with Cr-3$d$, S-3$p$ and O-3$p$ atomic orbitals [49]. The resulting Hamiltonian was symmetrized using WannSymm code [50]. To properly describe the on-site Coulomb interaction of the localized Cr-$d$ electron, we employed the LDA+$U$ approach with an additional Hubbard-like effective interaction with $U = 0$, 2, and 4 eV. For the high-temperature phase, the system remains metallic regardless of the $U$ value. In contrast, for the low-temperature phase, a band gap opens when $U > 2$ eV, and the system becomes insulating. When $U = 2.5$ eV, the calculated band gap for the low-temperature phase is approximately 200 meV, which is close to the experimental value.

## Acknowledgements

We thank Z. Ren for the assistance with the electrical transport and specific heat measurements. Y. Liu thanks the National Science Foundation of China (Grant No. 12474132 and 12004337) for support. G.-H. Cao acknowledges supports from the National Key Research and Development Program of China (Grant Nos. 2022YFA1403202, 2023YFA1406101). C.-C. Xu thanks the support from National Science Foundation of China (Grant No. 12304175). J.-K. B. acknowledges support from the Beijing National Laboratory for Condensed Matter Physics (Grant No. 2023BNLCMPKF019).

## Author Contributions

Y. Liu, C.-C Xu, J.-K Bao, and B.-J. Lv contributed equally to this work. Y. Liu synthesized the sample and performed the fundamental characterizations with the help of Y.-M. Lu, X.-Y. Zhao, X.-Y. Chen, H.-X. Li, W.-Z Yang and B.-R. Zhu. J.-K. Bao, J.-Y. Liu and Z.-Y. Zhang contributed to the structural analysis. Neutron diffraction experiments were conducted by B.-J. Lv and H. Li. Y. Liu contributed the magnetic structure analysis with the help from Y.-Q. Lin. C.-C. Xu and J. Li performed the theoretical calculations. The manuscript was written by Y. Liu, G.-H. Cao, J.-K. Bao, and C.-C. Xu, with input from all authors. Y. Liu and G.-H. Cao co-conceived the project.

## Conflict of Interest

The authors declare no conflict of interest.

## Data Availability Statement

Reasonable request of the data that support the findings of this study should be addressed to liuyiphy@zjut.edu.cn.

# Supporting information

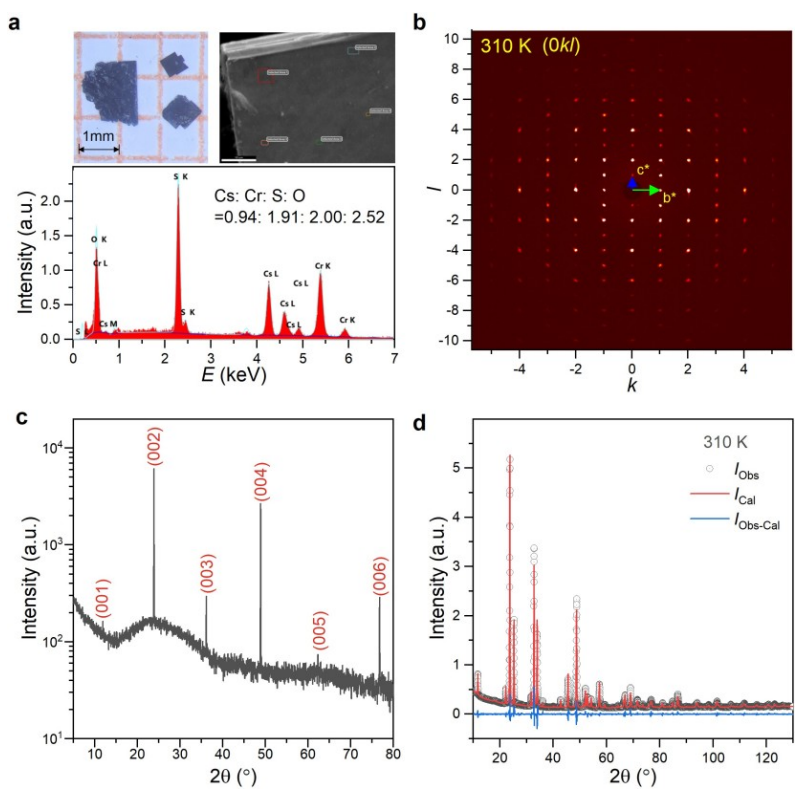

Figure S1: Characterizations of chemical compositions and crystal structure of $CsCr_2S_2O$. a, Optical and scanning electron microscopy images and the EDX spectrum. b, Reconstructed single-crystal XRD reciprocal space map of $(0kl)$ plane at 310 K. c, d, X-ray diffraction patterns for single crystals and polycrystals. The polycrystal diffraction pattern was refined using the $P4/mmm$ structural model. The observed ($I_{obs}$) and calculated ($I_{cal}$) profiles, along with their difference ($I_{obs} - I_{cal}$), are presented.

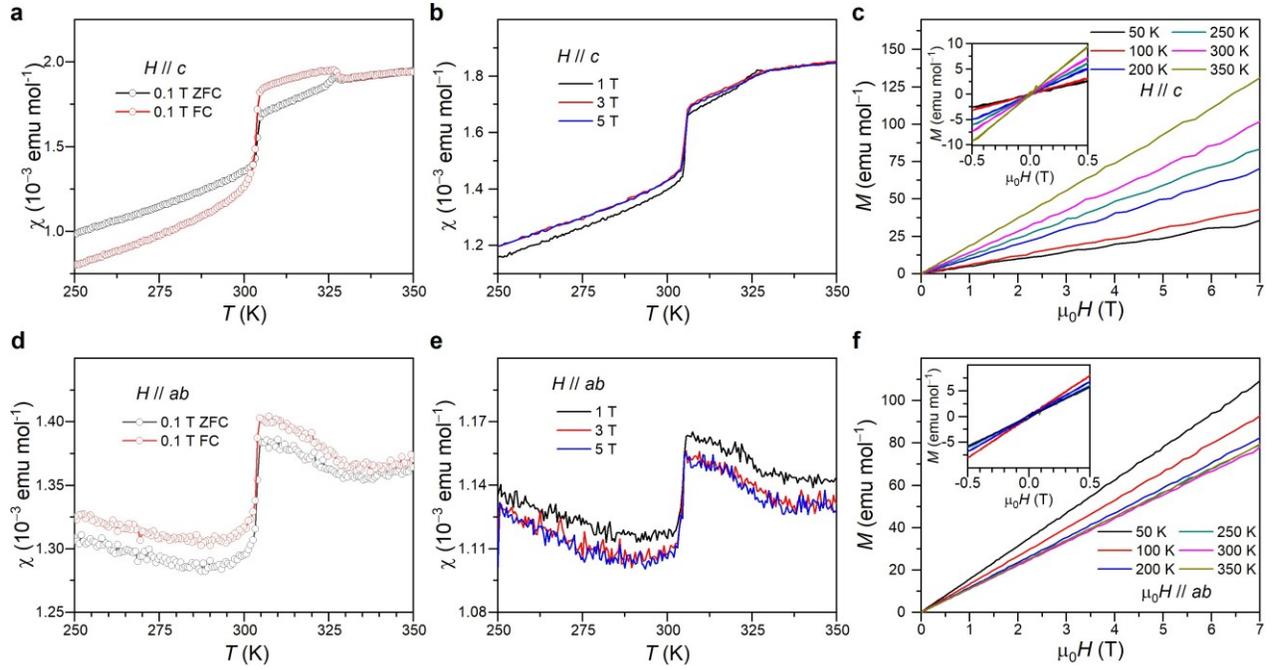

Figure S2: Anisotropic magnetic properties of $CsCr_2S_2O$. a, d, Anisotropic susceptibility in zero-field-cooling(ZFC) and field-cooling(FC) protocols under magnetic field of 0.1 T. b, e, Anisotropic susceptibility under higher magnetic fields. c, f, Anisotropic isothermal magnetization at vaious temperatures. The insets are magnetic hysteresis loop between -0.5 T and 0.5 T.

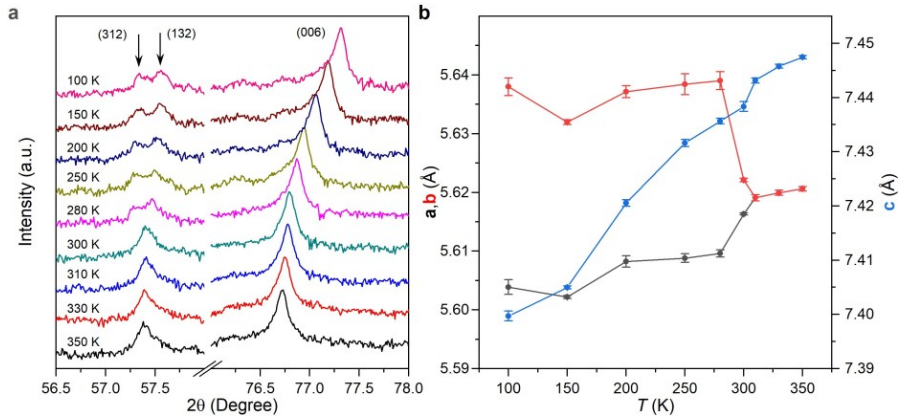

Figure S3: Powder x-ray diffractions of $CsCr_2S_2O$. a, The diffraction patterns focusing on (132) and (006) peaks under different temperatures. b, The temperature dependence of cell parameters obtained via Jade software. For comparison, the cell parameters $a$ and $b$ for the HT phase are multiplied by $\sqrt{2}$.

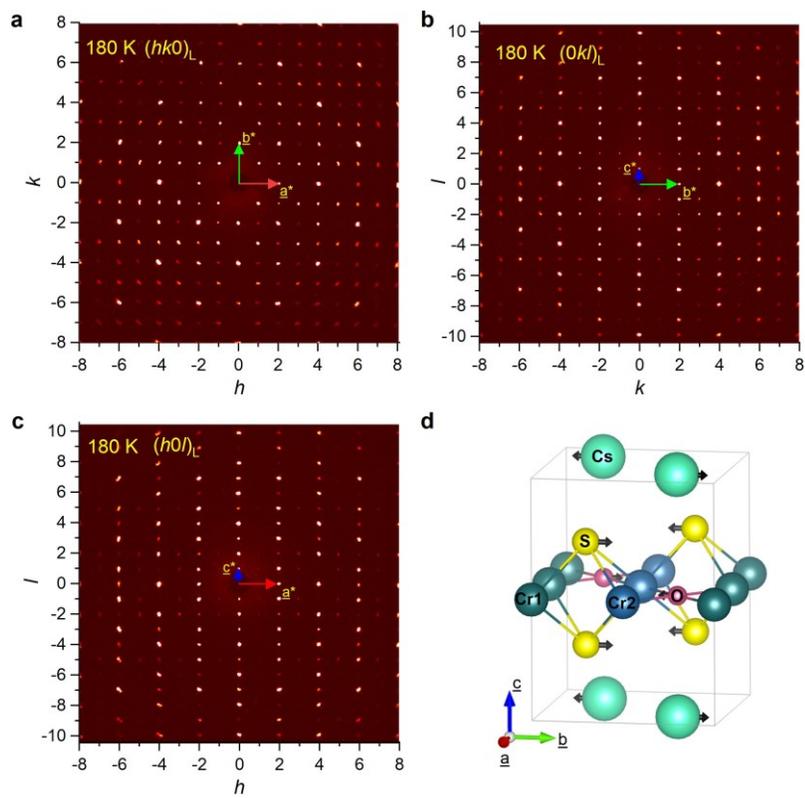

Figure S4: Low-temperature crystal structure $CsCr_2S_2O$. a, b, c, Reconstructed ($hk0$), ($0hk$) and ($h0l$) planes of reflections at 180 K indexed in space group *Pmam*. d, The crystal structure at 180 K. The arrow lengths represent the magnitude of atomic displacements from the center positions.

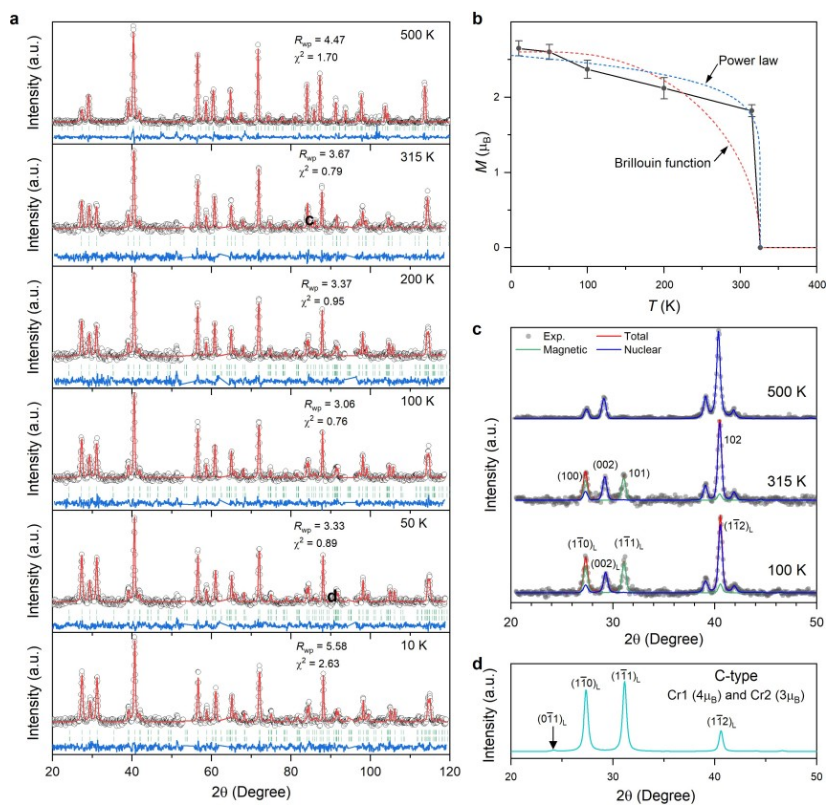

Figure S5: Characterization of the magnetic structure in CsCr$_2$S$_2$O. a, The NPD diffraction patterns from 500 K to 10 K, refined with both nuclear and magnetic contributions. b, Temperature dependence of the refined Cr magnetic moments. The blue and red dashed lines represent fits to power law and Brillouin function, respectively. c, The refinements with contributions from nuclear and magnetism. d, Simulation of magnetic peaks in LT phase with modulated magnetic moments at Cr1(4 $\mu_B$) and Cr2(3 $\mu_B$).

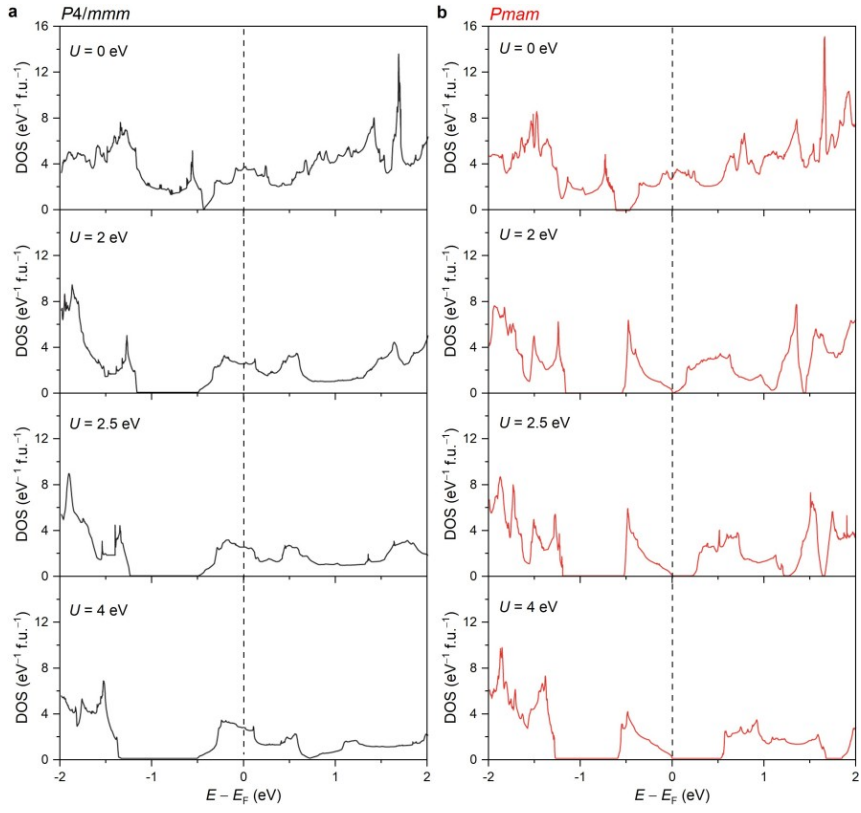

Figure S6: Dependence of the electronic density of states (DOS) on the Coulomb repulsion $U$ in $CsCr_2S_2O$ for AFM HT and LT phases. a,b, Evolution of DOS in the HT(left) and LT(right) phases.

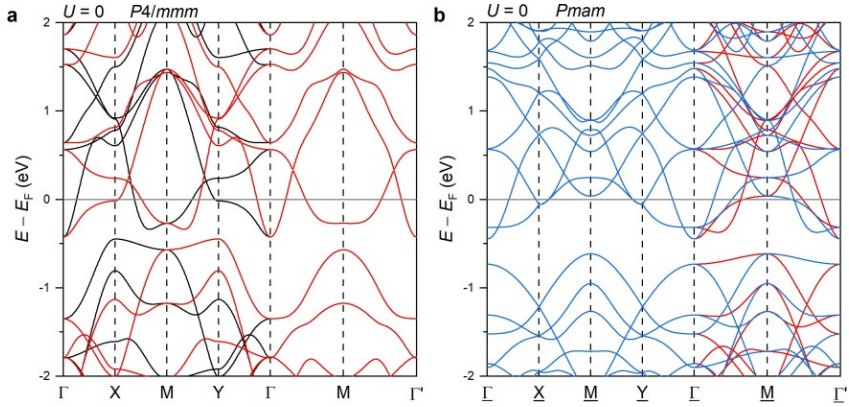

Figure S7: Band structures of $CsCr_2S_2O$ without $U$. a,b, Band structures with spin splitting at $U = 0$ for the HT(left) and LT(right) phases.

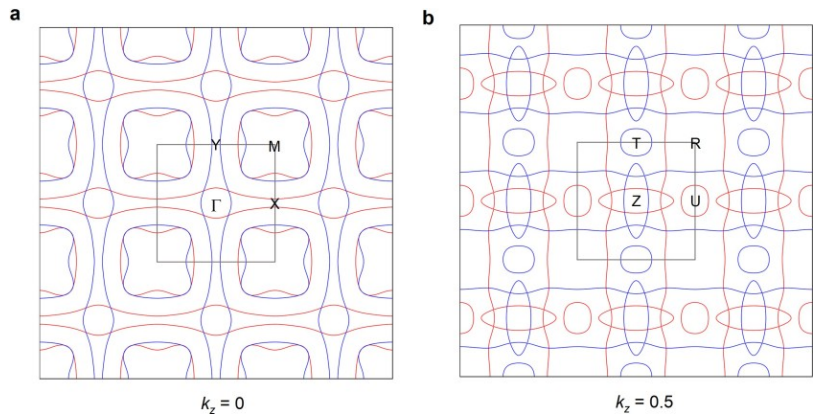

Figure S8: Calculated spin-resolved Fermi surface at $k_z = 0$ and $k_z = 0.5$. Red and blue lines correspond to spin-up and spin-down electrons, respectively.

**Table S1. Crystallographic data of CsCr$_2$S$_2$O at 310 K from the single-crystal XRD.**

| | |
|---|---|
| Empirical formula | CsCr$_2$S$_2$O |
| Formula weight | 317.03 |
| Temperature | 310 K |
| Wavelength | 0.71073 Å |
| Crystal system | Tetragonal |
| Space group | P4/mmm |
| Unit cell dimensions | $a$ = 3.9733(1) Å, α = 90° |
| | $b$ = 3.9733(1) Å, β = 90° |
| | $c$ = 7.4429(5) Å, γ = 90° |
| Volume | 117.502(10) Å$^3$ |
| Z | 1 |
| Density (calculated) | 4.480 g/cm$^3$ |
| Absorption coefficient | 12.967 mm$^{-1}$ |
| F(000) | 143 |
| Crystal size | 0.145 × 0.135 × 0.026 mm$^3$ |
| θ range for data collection | 2.737 to 30.459° |
| Index ranges | -5<=$h$<=5, -5<=$k$<=5, -10<=$l$<=10 |
| Reflections collected | 3971 |
| Independent reflections | 142 [$R_{int}$ = 0.0330] |
| Completeness to θ = 25.242° | 98.9% |
| Refinement method | Full-matrix least-squares on $F^2$ |
| Data / restraints / parameters | 142 / 0 / 11 |
| Goodness-of-fit | 0.505 |
| Final $R$ indices [$I > 2\sigma(I)$] | $R_{obs}$ = 0.0136, $wR_{obs}$ = 0.0499 |
| $R$ indices [all data] | $R_{all}$ = 0.0143, $wR_{all}$ = 0.0518 |
| Extinction coefficient | NA |
| Largest diff. peak and hole | 0.752 and -0.410 e·Å$^{-3}$ |

$R = \Sigma||F_o|-|F_c||/\Sigma|F_o|$, $wR = \{\Sigma[w(|F_o|^2 - |F_c|^2)^2]/\Sigma[w(|F_o|^4)]\}^{1/2}$ and $w=1/[\sigma^2(F_o^2)+(0.1000P)^2]$ where $P=(F_o^2+2F_c^2)/3$

| Label | x | y | z | Occupancy | $U_{eq}$* | $U_{11}$ | $U_{22}$ | $U_{33}$ |
|---|---|---|---|---|---|---|---|---|
| Cs1 | 0 | 0 | 0 | 1 | 20(1) | 17(1) | 17(1) | 25(1) |
| Cr1 | 0.5 | 0 | 0.5 | 1 | 11(1) | 5(1) | 8(1) | 21(1) |
| S1 | 0.5 | 0.5 | 0.7115(2) | 1 | 17(1) | 19(1) | 19(1) | 15(1) |
| O1 | 0 | 0 | 0.5 | 1 | 13(1) | 3(1) | 3(1) | 35(2) |

*$U_{eq}$ is defined as one third of the trace of the orthogonalized $U_{ij}$ tensor. The anisotropic displacement factor exponent takes the form: $-2\pi^2[h^2a^{*2}U_{11} + ... + 2hka^*b^*U_{12}]$. $U_{12}$, $U_{13}$ and $U_{23}$ are zero. The unit of the equivalent isotropic and anisotropic displacement parameters is 0.001 Å$^2$.

| Bond | Distances(Å) |
|---|---|
| Cr1-O1 | 1.9867(1) |
| Cr1-S1 | 2.5345(8) |
| Cr1-Cr1 | 2.8096(1) |

**Table S2. Crystallographic data of CsCr$_2$S$_2$O at 180 K from the single-crystal XRD.**

| | |
|---|---|
| Empirical formula | CsCr$_2$S$_2$O |
| Formula weight | 317.03 |
| Temperature | 180 K |
| Wavelength | 0.71073 Å |
| Crystal system | Orthorhombic |
| Space group | *Pmam* |
| Unit cell dimensions | $a$ = 5.6049(2) Å, α = 90° |
| | $b$ = 5.6285(2) Å, β = 90° |
| | $c$ = 7.4054(3) Å, γ = 90° |
| Volume | 233.619(15) Å$^3$ |
| Z | 1 |
| Density (calculated) | 4.5066 g/cm$^3$ |
| Absorption coefficient | 13.044 mm$^{-1}$ |
| F(000) | 286 |
| Crystal size | 0.145 x 0.135 x 0.026 mm$^3$ |
| θ range for data collection | 2.75 to 72.91° |
| Index ranges | -15<=$h$<=13, -15<=$k$<=13, -19<=$l$<=19 |
| Reflections collected | 33020 |
| Independent reflections | 2643 [$R_{int}$ = 0.0609] |
| Completeness to θ = 72.5° | 98% |
| Refinement method | Full-matrix least-squares on $F^2$ |
| Data / restraints / parameters | 2643 / 0 / 24 |
| Goodness-of-fit | 1.5737 |
| Final R indices [$I > 3σ(I)$] | $R_{obs}$ = 0.0318, $wR_{obs}$ = 0.0387 |
| R indices [all data] | $R_{all}$ = 0.0382, $wR_{all}$ = 0.0398 |
| Extinction coefficient | NA |
| Largest diff. peak and hole | 5.30 and -1.39 e·Å$^{-3}$ |

$R = Σ||F_o|-|F_c|| / Σ|F_o|$, $wR = \{Σ[w(|F_o|^2 - |F_c|^2)^2] / Σ[w(|F_o|^4)]\}^{1/2}$ and $w=1/[σ^2(F_o^2)+(0.1000P)^2]$ where $P=(F_o^2+2F_c^2)/3$

| Label | x | y | z | Occupancy | $U_{eq}$* | $U_{11}$ | $U_{22}$ | $U_{33}$ |
|---|---|---|---|---|---|---|---|---|
| Cs1 | 0.25 | 0.2398(1) | 0 | 1 | 11(1) | 10(1) | 10(1) | 13(1) |
| Cr1 | 0 | 0 | 0.5 | 1 | 9(1) | 5(1) | 4(1) | 18(1) |
| Cr2 | 0 | 0.5 | 0.5 | 1 | 6(1) | 4(1) | 4(1) | 9(1) |
| O1 | 0.25 | 0.2510(2) | 0.5 | 1 | 10(1) | 5(1) | 4(1) | 20(1) |
| S1 | 0.75 | 0.2745(1) | 0.7129(1) | 1 | 8(1) | 7(1) | 9(1) | 8(1) |

*$U_{eq}$ is defined as one third of the trace of the orthogonalized $U_{ij}$ tensor. The anisotropic displacement factor exponent takes the form: $-2π^2[h^2a^{*2}U_{11} + ... + 2hka^*b^*U_{12}]$. $U_{12}$, $U_{13}$ and $U_{23}$ are zero. The unit of the equivalent isotropic and anisotropic displacement parameters is 0.001 Å$^2$.

| Bond | Distances(Å) |
|---|---|
| Cs1-Cs1 | 4.0519(2) × 4 |
| Cr1-S1 | 2.6148(3) × 4 |
| Cr1-O1 | 1.9896(8) × 2 |
| Cr1-Cr1 | 2.8024(2) × 2 |
| Cr1-Cr2 | 2.8142(2) × 2 |
| Cr2-Cr2 | 2.8024(2) × 2 |
| Cr2-S1 | 2.4618(3) × 4 |
| Cr2-O1 | 1.9820(8) × 2 |